\documentclass[aps, prb, preprint, superscriptaddress,preprintnumbers,amssymb]{revtex4-1}
\usepackage{graphicx}
\usepackage{dcolumn}
\usepackage{bm}

\begin{document}

\preprint{ver4p2}

\title{In-plane and Interlayer Magnetoresistances in FeSe}



\author{Taichi Terashima}
\email{TERASHIMA.Taichi@nims.go.jp}
\affiliation{Research Center for Materials Nanoarchitectonics (MANA), National Institute for Materials Science, Tsukuba 305-0003, Japan}
\author{Shinya Uji}
\affiliation{Research Center for Materials Nanoarchitectonics (MANA), National Institute for Materials Science, Tsukuba 305-0003, Japan}
\author{Hiroaki Ikeda}
\affiliation{Department of Physics, Ritsumeikan University, Kusatsu, Shiga 525-8577, Japan}
\author{Yuji Matsuda}
\altaffiliation[Present address: ]{MPA-Q, Los Alamos National Laboratory, Los Alamos, NM 87545, USA}
\affiliation{Department of Physics, Kyoto University, Kyoto 606-8502, Japan}
\author{Takasada Shibauchi}
\affiliation{Department of Advanced Materials Science, University of Tokyo, Kashiwa, Chiba 277-8561, Japan}
\author{Shigeru Kasahara}
\email{kasa@okayama-u.ac.jp}
\affiliation{Research Institute for Interdisciplinary Science, Okayama University, Okayama 700-8530, Japan}


\date{\today}

\begin{abstract}
We report measurements of the in-plane and interlayer magnetoresistances of FeSe.
The in-plane magnetoresistance $\Delta \rho_{ab}/\rho_{ab}(0)$ for $B \parallel c$ is positive below $T_s$ and increases with decreasing temperature, exceeding 2.5 at $T$ = 10 K and $B$ = 14 T.
The field-direction dependence indicates that the in-plane magnetoresistance is basically determined by the $c$-axis component of the magnetic field.
The interlayer magnetoresistance $\Delta \rho_{c}/\rho_{c}(0)$ is negative below $T_s$ but turns positive below $\sim$18 K, which is probably due to the contamination by the large in-plane magnetoresistance.
The field-direction dependence of the interlayer magnetoresistance can approximately be described by a standard formula for quasi-two-dimensional electron systems except near $B \parallel ab$.
The experimental magnetoresistance near $B \parallel ab$ is larger than the formula, which can be attributed to the so-called interlayer coherence peak.
The large width of the peak indicates the correspondingly large interlayer transfer energy.
\end{abstract}


\maketitle



%

\section*{1. I\lowercase{ntroduction}}

Magnetoresistance measurements have a long history as a technique for studying the electronic state of metals \cite{PippardRPP1960} and are still actively used today. 
In such measurements, the mutual orientation of electrical current and magnetic field is important. 
By measuring magnetoresistance as a function of magnetic field orientation, one can investigate the details of the Fermi surface. 
In a well-known example, the direction of open orbits can be determined by such a measurement. 
In this study, we measure the magnetoresistance effects on the in-plane and interlayer resistivites of the iron-based superconductor FeSe with varying magnetic field directions, and we reveal the quasi-two-dimensional nature of the electronic structure of FeSe.

FeSe is an intriguing iron-based superconductor parent compound. \cite{Hsu08PNAS}
Unlike typical parent compounds, FeSe exhibits a structural phase transition associated with electronic nematic ordering at $T_s \sim90$ K, but not antiferromagnetic ordering. 
Furthermore, it becomes superconducting below $T_c \sim9$ K.
It is argued that this superconductivity is close to the BCS--BEC (Bardeen--Cooper--Schrieffer--Bose--Einstein condensate) cross-over regime.\cite{Kasahara14PNAS}
Quantum oscillation \cite{Terashima14PRB, Audouard15EPL, Watson15PRB} and angle-resolved photoemission spectroscopy studies \cite{Tan13NatMat, Nakayama14PRL, Shimojima14PRB, Maletz14PRB, Watson15PRB} have shown that the electronic structure in the electronic nematic phase differs significantly from that predicted by density functional theory. 
The Fermi surface calculated by the density functional theory includes three hole cylinders and two electron cylinders, but the experimental studies indicate that there are only one hole and one electron cylinder. \cite{Terashima14PRB, Yi19PRX}
The failure of the theory renders the experimental characterization of the electronic structure in FeSe vital.
In this study, we confirm that the Fermi surface in the nematic phase consists of quasi-two-dimensional modulated cylinders by analyzing the magnetic-field direction dependence of magnetoresistance.

\section*{2. E\lowercase{xperimental} R\lowercase{esults} \lowercase{and} D\lowercase{iscussion}}
High-quality single crystals of FeSe were grown by a chemical vapor transport method. \cite{Bohmer13PRB}
Electrical contacts were spot-welded and reinforced with conducting silver paint.
A current contact and a voltage contact were attached to each (001) plane of a sample for interlayer resistivity measurements, whereas four contacts were attached to the same (001) plane for in-plane resistivity measurements as usual. 
Because samples were not detwinned, they were twinned below $T_s$, which basically obscured in-plane anisotropy arising from the orthorhombicity in the nematic phase.
Samples were mounted on a two-axis rotation platform to enable the control of both the polar $\theta$ and azimuthal $\phi$ angles of the magnetic field.
Resistivity measurements were performed in a 17-T superconducting magnet and a $^4$He variable temperature insert.

\begin{figure}
\includegraphics[width=8.6cm]{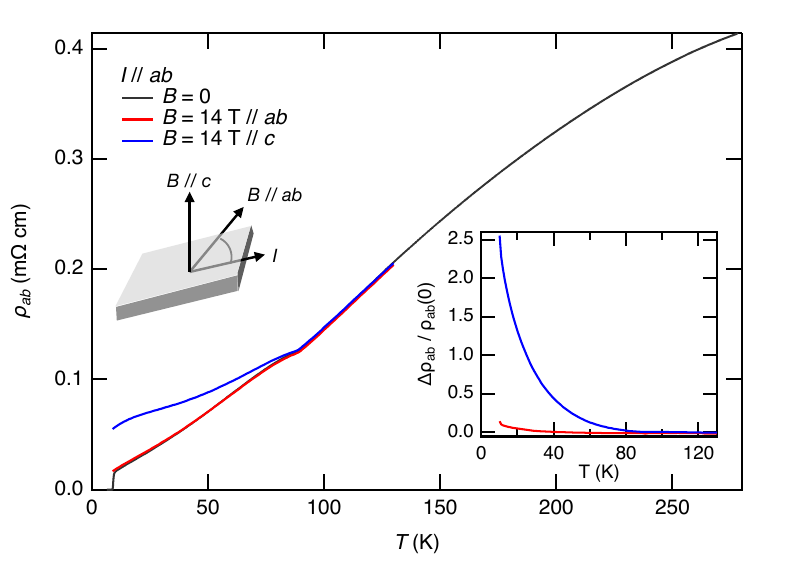}
\caption{\label{Sample}(Color online) Temperature dependence of in-plane resistivity $\rho_{ab}$  in FeSe sample 1 ($0.64 \times 0.8 \times 0.02$~mm$^3$).
The data for $B$ = 0 and $B$ = 14 T applied parallel to the $ab$ plane and along the $c$ axis are shown.
The left inset shows the geometry of the current $I$ and the magnetic field $B$ schematically.
The in-plane field is approximately at a 50$^{\circ}$ angle from the current.
The right inset shows the in-plane magnetoresistance $\Delta \rho_{ab} / \rho_{ab}(0)$ at $B$ = 14 T for $B \parallel ab$ and $B \parallel c$.
}
\end{figure}

\begin{figure}
\includegraphics[width=8cm]{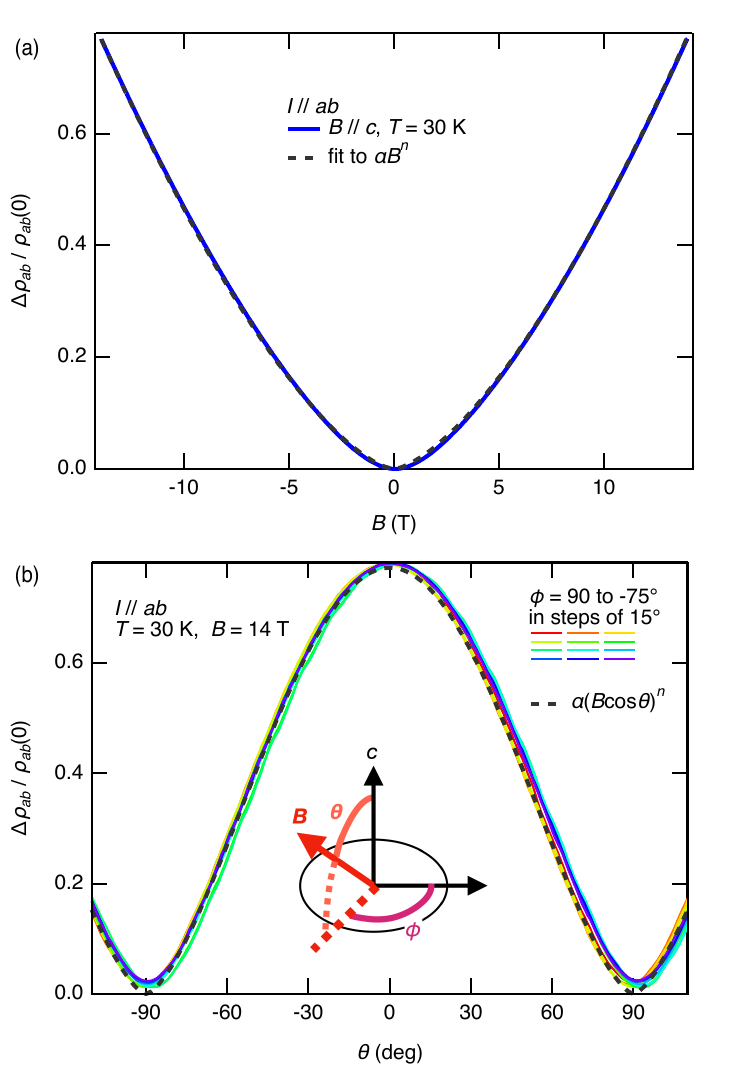}
\caption{\label{Sample}(Color online) Magnetoresistance effects on in-plane resistivity in FeSe sample 1.
(a) In-plane magnetoresistance $\Delta \rho_{ab} / \rho_{ab}(0)$ as a function of magnetic field along the $c$ axis measured at $T$ = 30 K.
A fit to $\alpha B^n$ (broken line) gives $\alpha$ = 0.01433(2) and $n$ = 1.512(1).
(b) In-plane magnetoresistance at $T$ = 30 K and $B$ = 14 T as a function of $\theta$, which is the polar angle of the magnetic field direction measured from the $c$ axis.
The azimuth angle $\phi$ specifies the rotation plane of the magnetic field and was varied from $\phi$ = 90 to -75$^{\circ}$ in steps of 15$^{\circ}$.
The broken line shows $\alpha (B \cos \theta) ^n$ with the same values of $\alpha$ and $n$ as in (a).
The inset explains the field angles $\theta$ and $\phi$.
The origin of $\phi$ is defined with respect to the sample holder, not to a crystal axis.
}
\end{figure}

We begin with in-plane resistivity.
Figure 1 shows the temperature dependence of the in-plane resistivity in sample 1 at zero field and at a field of 14 T applied along the $ab$ plane and the $c$ axis.
The current was parallel to the tetragonal [100] direction.
The $ab$-plane field was at an angle of about 50$^{\circ}$ to the current [$\phi$ = -15$^{\circ}$ in Fig. 2(b)].
Although this configuration is neither a transverse one nor a longitudinal one, the in-plane angle between the current and the field is not very important, as we will see in Fig. 2(b).
The measurements under magnetic field were performed at both the positive field ($B$ = 14 T) and the negative field ($B$ = $-$14 T), and the measured voltages were symmetrized, i.e., $V_{\mathrm{sym}}=[V(14 \, \mathrm{T}) +V(-14 \, \mathrm{T})]/2$, to remove the contamination by the Hall voltage, although the contamination was found to be negligibly small, as we will see in Fig. 2(a).
The right inset of Fig. 1 shows the magnetoresistance $\Delta\rho_{ab}/\rho_{ab}(0)$, where  $\Delta\rho_{ab}$ is the difference between the resistivity at $B$ = 0 and 14 T and $\rho_{ab}(0)$ the resistivity at $B$ = 0. 
The magnetoresistance is almost negligible above $T_s$ (= 89.0 K for this sample) for both field directions, $B \parallel ab$ and $B \parallel c$.
For $B \parallel c$, the magnetoresistance increases rapidly with decreasing temperature below $T_s$, exceeding 2.5 at $T$ = 10 K.
For $B \parallel ab$, the magnetoresistance is small even below $T_s$, being 0.15 at $T$ = 10 K.
The observed behavior is qualitatively consistent with those described in previous reports, \cite{KnonerPRB15, Amigo24PRB} although the magnitude of the magnetoresistance for $B \parallel c$ is larger in the present case.

Figure 2(a) shows the magnetic field dependence of the in-plane magnetoresistance for $B \parallel c$ measured at $T$ = 30 K.
The nearly perfect symmetry of the data with respect to $B$ = 0 confirms that the Hall-voltage contamination is negligible.
A fit to $\alpha B^n$ (broken line) gives $n$ = 1.512(1).
Although this exponent is smaller than $n$ = 2 expected from a simple two-carrier model of compensated metals, a similar subquadratic magnetoresistance is often observed in nominally compensated semimetals.
On the basis of a detailed analysis of magnetoresistance in a compensated semimetal antimony, it has been argued in Ref. \onlinecite{Benoit18PRM} that the subquadratic magnetoresistance can be attributed to imperfect compensation and field-dependent mobility.

The magnitude and field dependences of the in-plane magnetoresistance are broadly consistent with our previous reports. \cite{Terashima16PRB, Terashima16PRB2}
It has been reported previously that the magnetoresistance of FeSe does not follow Kohler's rule, \cite{Rossler15PRB, Terashima16PRB2} but in this study, we do not discuss this point owing to insufficient data collected on the matter 

Figure 2(b) shows the $\theta$ dependence of the magnetoresistance at $B$ = 14 T and $T$ = 30 K, where the polar angle $\theta$ of the field direction was measured from the $c$ axis.
The rotation plane of the magnetic field was specified by the azimuth angle $\phi$, and $\phi$ = 35$^{\circ}$ approximately corresponds to the tetragonal (010) plane, which is parallel to the current.
Figure 2(b) shows the data obtained for different $\phi$ values together, and they overlap each other.
The broken line shows $\alpha (B\cos \theta)^n$ calculated with the same values of $\alpha$ and $n$ as those in Fig. 2(a).
It matches the experimental curves.
This indicates that the in-plane magnetoresistance is basically determined by the $c$-axis component of the magnetic field ($B\cos \theta$), as expected for quasi-two-dimensional electron systems.
In Appendix A, we show the data presented in Figs. 2(a) and 2(b) together as a function of $B\cos\theta$.

\begin{figure}
\includegraphics[width=8.6cm]{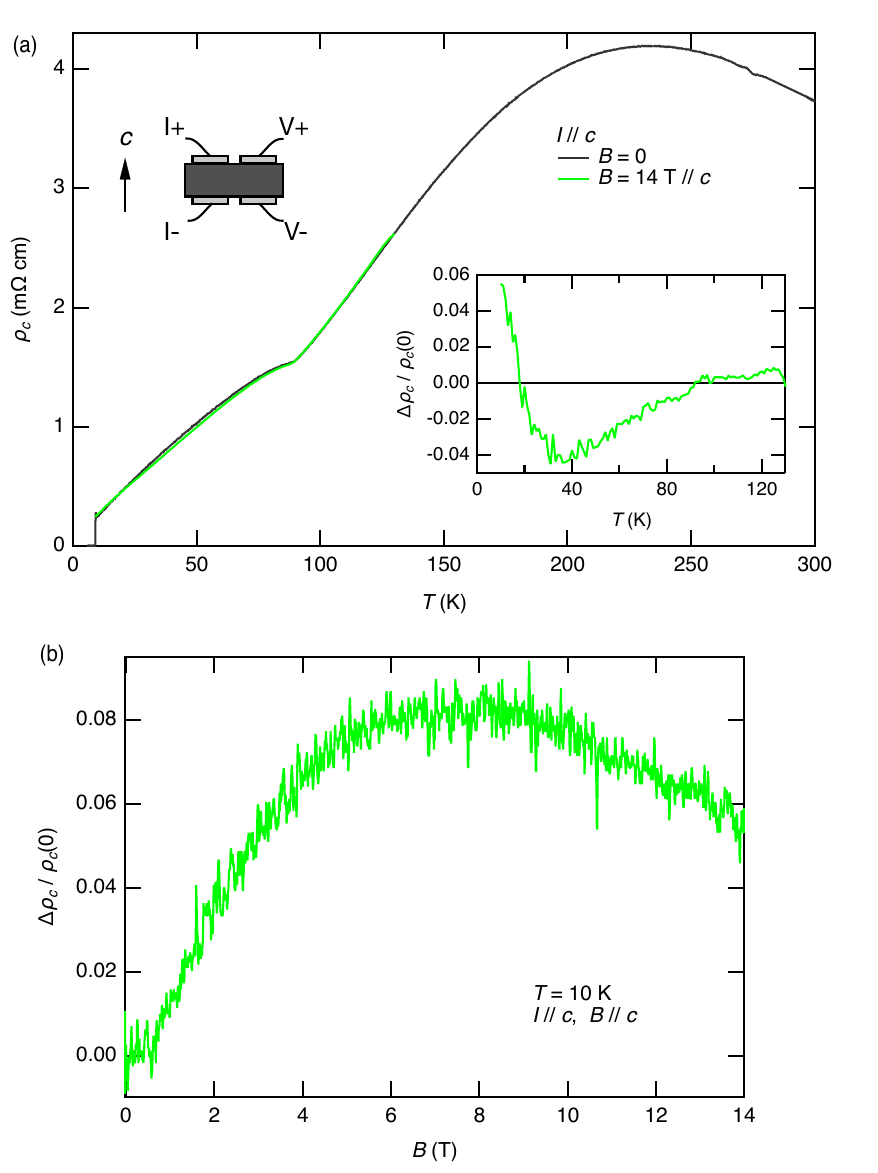}
\caption{\label{Sample}(Color online) Temperature and magnetic-field dependences of interlayer resistivity in FeSe sample 2 ($1.02 \times 0.6 \times 0.135$~mm$^3$).
(a) Temperature dependence of interlayer resistivity measured at $B$ = 0 and 14 T applied parallel to the $c$ axis.
The lower right inset shows the interlayer magnetoresistance $\Delta \rho_{c} / \rho_{c}(0)$ at $B$ = 14 T parallel to $c$.
The upper left inset is a schematic of the contact arrangement.
(b) Interlayer magnetoresistance as a function of magnetic field along the $c$ axis measured at $T$ = 10 K.
}
\end{figure}

\begin{figure}
\includegraphics[width=8.6cm]{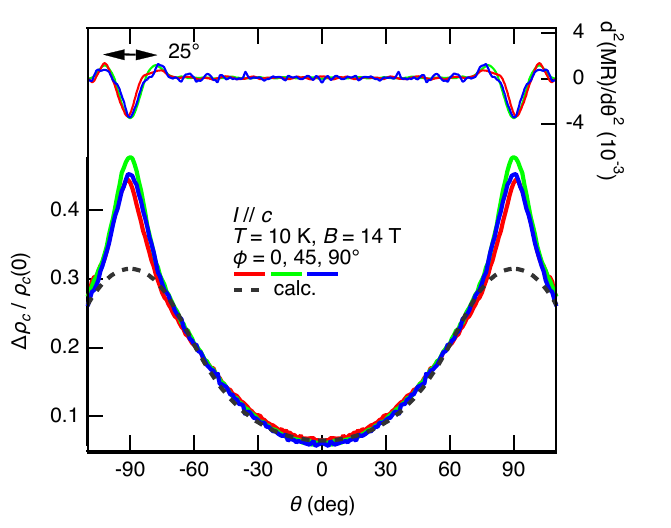}
\caption{\label{Sample}(Color online) Interlayer magnetoresistance in FeSe sample 2 at $T$ = 10 K and $B$ = 14 T as a function of the field angle $\theta$.
The upper curves are the corresponding second derivatives with respect to $\theta$ [MR means $\Delta\rho_c/\rho_c(0)$].
Three field rotation planes $\phi$ = 0, 45, and 90$^{\circ}$ were used.
The broken line was calculated using Eq. (1) with $c k_{\mathrm{F}}$ = 0.5 and $\omega_0 \tau$ = 1.5.
}
\end{figure}

We now switch to interlayer resistivity.
Figure 3(a) shows the temperature dependence of interlayer resistivity in sample 2 at zero field and at $B$ = 14 T parallel to the $c$ axis.
For $B$ = 14 T, symmetrized data are shown.
A schematic of the contact arrangement is shown in the upper left inset.
The interlayer resistivity exhibits a maximum near 230 K, which is consistent with a previous report by Amig\'o \textit{et al.}, \cite{Amigo24PRB} but note that the present peak temperature $T^{\mathrm{max}}$ = 233 K is slightly higher than 229 K reported in Ref. \onlinecite{Amigo24PRB}. 
The lower right inset shows the interlayer magnetoresistance $\Delta\rho_c/\rho_c(0)$ at $B$ = 14 T parallel to $c$, which becomes negative below $T_s$, showing a negative peak at around 35 K, and turns positive below $\sim$18 K.
The magnitude of the interlayer magnetoresistance is much smaller than that of the in-plane one.
Figure 3(b) shows the interlayer magnetoresistance $\Delta\rho_c/\rho_c(0)$ at $T$ = 10 K as a function of the field applied along the $c$ axis.
Although it is positive at low fields, a negative component appears above about 8 T.

Amig\'o \textit{et al.} \cite{Amigo24PRB} reported that the interlayer magnetoresistance for $B \parallel c$ was negative below $T_s$ down to $T_c$.
The positive magnetoresistance that we observed below $\sim$18 K probably indicates that our measurements were contaminated by in-plane resistivity.
An experimental interlayer resistivity may be contaminated by in-plane resistivity:
for example, if cleavage occurs inside a crystal, the electrical current has to flow along the in-plane direction to avoid the cleavage, which makes the measured voltages contaminated  by an in-plane resistivity component.
Amig\'o \textit{et al.} pointed out that the peak temperature $T^{\mathrm{max}}$ was a good measure of the contamination by in-plane resistivity. \cite{Amigo24PRB}
Because the in-plane resistivity increases monotonically in a temperature range near $T^{\mathrm{max}}$, the apparent $T^{\mathrm{max}}$ shifts to a higher temperature as the contamination by in-plane resistivity increases.
Because the in-plane magnetoresistance for $B \parallel c$ is positive and large, especially at low temperatures (Fig. 1), the apparent interlayer magnetoresistance becomes positive as the amount of contamination increases.
The fact that the present $T^{\mathrm{max}}$ = 233 K is  larger than 229 K in Ref. \onlinecite{Amigo24PRB} suggests that our measurements were slightly more contaminated by the in-plane resistivity component, which explains the present observation of the positive interlayer magnetoresistance for $B \parallel c$ at low temperatures.
In Appendix B, we show the results of `interlayer resistivity' measurements on other samples, in which $T^{\mathrm{max}}$ is still higher, and the interlayer magnetoresistance for $B \parallel c$ appears positive and large because of the in-plane resistivity contamination.

Figure 4 shows the interlayer magnetoresistance at $B$ = 14 T and $T$ = 10 K as a function of the polar angle $\theta$ of the field direction.
The data are symmetrized with respect to $\theta$ = 0.
Three field-rotation planes, $\phi$ = 0, 45, and 90$^{\circ}$, were used, where $\phi$ = 45$^{\circ}$ approximately corresponds to the tetragonal (100) plane.
Although the three curves do not match perfectly, which may partly be ascribed to the misalignment of the sample, the interlayer magnetoresistance is almost independent of $\phi$.
This is not surprising because the sample was not detwinned.
The field-angle dependence is quite different from that of the in-plane magnetoresistance in Fig. 2(b).
The magnetoresistance is the largest at $\theta$ = $\pm$90$^{\circ}$, where the field is perpendicular to the current.
This is reasonable because the transverse magnetoresistance ($B \bot I \parallel c$) is usually larger than the longitudinal one ($B \parallel I \parallel c$).
There is a finite magnetoresistance $\Delta\rho_c/\rho_c(0)$ = 0.065 at $\theta$ = 0, which might indicate a slight contamination by in-plane resistivity.
However, the magnitude is much smaller than that observed in the in-plane resistivity measurements shown in Fig. 1, the right inset of which indicates that the in-plane magnetoresistance amounts to 2.6 at $T$ = 10 K.
The considerably smaller magnitude warrants that the contamination by in-plane resistivity is limited and that the field-angle dependence in Fig. 4 captures that of the interlayer magnetoresistance with sufficient accuracy.

For a quasi-two-dimensional metal with a weak $c$-axis energy dispersion of the form $\cos(c k_z)$, the interlayer conductivity under magnetic field (except near $\theta$ = $\pm$90$^{\circ}$) is given by \cite{Yagi90JPSJ}
\begin{equation}
\sigma_{zz} = \sigma_{zz}^0 \left[J_0^2\left(c k_{\mathrm{F}} \tan \theta\right) + \sum_{\nu = 1}^{\infty} \frac{2J_{\nu}^2 \left(c k_{\mathrm{F}} \tan \theta \right)}{1+\left(\omega_0 \tau \nu \cos \theta \right)^2} \right] ,
\end{equation}
where $\sigma_{zz}^0$ is the interlayer conductivity  at zero magnetic field, $J_{\nu}$ the $\nu$-th order Bessel function, $k_{\mathrm{F}}$ the in-plane Fermi wave vector, $\omega_0 = e/Bm^*$ the cyclotron frequency for $\theta$ = 0, $m^*$ the effective mass, and $\tau$ the relaxation time. \footnote{Eq. (1) is usually regarded as an approximation that is valid when $t_c / E_F \ll 1$.  However, the required condition is $(t_c / E_F)c k_F \ll 1$. Because $c k_F \approx 0.5$ in FeSe, the condition for $t_c / E_F$ is less strict.}
The quantum-oscillation data in Ref. \onlinecite{Terashima14PRB} indicate that $c k_{\mathrm{F}}$ = 0.47 and 0.60 for two Fermi cylinders of FeSe (i.e., electron and hole).
Accordingly, we set $c k_{\mathrm{F}}$ = 0.5 and calculated the interlayer magnetoresistance $\Delta \rho_c / \rho_c(0) = \sigma_{zz}^0 / \sigma_{zz} - 1$ at various $\omega_0 \tau$ values.
The broken line in Fig. 4 was calculated with $\omega_0 \tau$ = 1.5.
Because Eq. (1) gives $\Delta \rho_c / \rho_c(0)$ = 0 at $\theta$ = 0, we added a constant shift of 0.065 so that the calculated curve and the experimental result match at $\theta$ = 0.
The calculated curve reproduces the experimental result reasonably well except near $\theta$ = $\pm$90$^{\circ}$.
In Ref. \onlinecite{Terashima14PRB}, the mean free path of carriers was estimated for two quantum-oscillation frequencies, which corresponds to $\omega_0 \tau$ = 0.8 and 1.2 at $B$ = 14 T.
Hence, the above assumption of $\omega_0 \tau$ = 1.5 is reasonable.

The experimental magnetoresistance near $\theta$ = $\pm$90$^{\circ}$ is distinctly larger than the calculated one.
The enhancement of resistance becomes more evident in the second derivatives (upper curves).
This is due to a so-called interlayer coherence peak.
The approximations used to derive Eq. (1) do not hold near $\theta$ = $\pm$90$^{\circ}$, i.e., $B \parallel ab$, where the magnetoresistance is enhanced over Eq. (1) and peaks at $\theta$ = $\pm$90$^{\circ}$ because of small closed orbits \cite{Hanasaki98PRB} or self-crossing orbits \cite{Peschansky99PRB} formed on the sides of quasi-two-dimensional Fermi cylinders.
The appearance of such a peak is an indication of coherent interlayer transport, and hence, the peak was called the interlayer coherence peak. 
The width of the peak can be related to the magnitude of the interlayer transfer energy $t_c$.
For a single Fermi cylinder without in-plane anisotropy and with a $c$-axis dispersion of the form $\cos(c k_z)$, the relation is described by $\delta \theta \approx 2 c k_F t_c / E_F$. \cite{Hanasaki98PRB}
The interlayer coherence peak was initially found in organic conductors, \cite{Kartsovnik88JETPLett} but was also observed in the iron-based superconductor (parent) materials KFe$_2$As$_2$ \cite{Kimata10PRL} and CaFeAsF. \cite{Terashima22PRB}
In the present case, the width of the coherence peak, estimated from the second derivatives (upper curves in  Fig. 4), is about $\delta \theta \sim$ 25$^{\circ}$.
If we apply the above relation to this width, we obtain $t_c / E_F \approx$ 0.4.
However, this estimate should not be taken literally because the Fermi-surface model used to derive the relation is very much simplified, as described above.
Nonetheless the large magnitude of the interlayer transfer is reasonable because previous quantum-oscillation measurements indicated that the minimum and maximum cross-sectional areas of the Fermi cylinders considerably differ. \cite{Terashima14PRB}

The field-angle dependence of the interlayer magnetoresistance in FeSe and KFe$_2$As$_2$ is normal in the sense that the interlayer magnetoresistance is the smallest when $B \parallel c$, as expected from Eq. (1). \cite{Kimata10PRL}
On the other hand, it is unusual in CaFeAsF in that the interlayer magnetoresistance is the largest when $B \parallel c$, although the observed behavior was reproduced by a detailed calculation based on a first-principles electronic band structure. \cite{Terashima22PRB}
The distinct behavior may be related to the fact that the electronic structure in CaFeAsF is much more two-dimensional than those in FeSe and KFe$_2$As$_2$. \cite{Terashima22PRB}

In summary, we studied the in-plane and interlayer magnetoresistances in FeSe.
The in-plane magnetoresistance $\Delta \rho_{ab}/\rho_{ab}(0)$ for $B \parallel c$ was positive below $T_s$ and increased with decreasing temperature.
It exceeded 2.5 at $T$ = 10 K and $B$ = 14 T for the present sample.
Its dependence on the field direction indicated that it was basically determined by the $c$-axis component of the magnetic field, as is expected for quasi-two-dimensional electron systems.
The interlayer magnetoresistance $\Delta \rho_{c}/\rho_{c}(0)$ at $B$ = 14 T was initially negative below $T_s$ but turned positive below $\sim$18 K, which is probably due to the contamination by the large in-plane magnetoresistance.
The field-angle dependence of the interlayer magnetoresistance was reasonably well described by a standard formula for quasi-two-dimensional electron systems [Eq. (1)] with reasonable parameters $c k_F$ = 0.5 and $\omega_0 \tau$ = 1.5 except near $B \parallel ab$.
The experimental magnetoresistance near $B \parallel ab$ was larger than the magnetoresistance calculated with Eq. (1), which was attributed to the interlayer coherence peak.
The large width of the peak indicated the correspondingly large interlayer transfer, which is in agreement with previous quantum-oscillation results. \cite{Terashima14PRB}

\begin{acknowledgments}
This work was supported by Grants-in-Aid for Scientific Research on Innovative Areas ``Quantum Liquid Crystals'' (Nos. JP19H05824 and JP22H04485), Grants-in-Aid for Scientific Research(A) (Nos. JP21H04443, JP22H00105, and JP23H00089), a Grant-in-Aid for Scientific Research(B) (No. JP22H01173), a Grant-in-Aid for Scientific Research(C) (No. JP22K03537), and the Fund for the Promotion of Joint International Research (No. JP22KK0036) from Japan Society for the Promotion of Science.
MANA is supported by World Premier International Research Center Initiative (WPI), MEXT, Japan.
\end{acknowledgments}

\appendix

\section{$B\cos\theta$ Dependence}

\setcounter{figure}{0}
\renewcommand{\figurename}{FIG. A}

\begin{figure}
\includegraphics[width=7cm]{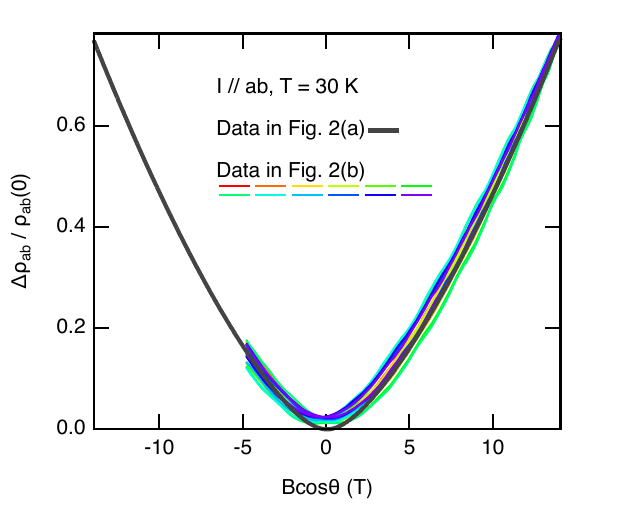}
\caption{\label{Sample}(Color online) In-plane magnetoresistance in FeSe sample 2 as a function of $B\cos\theta$.
The data in Figs. 2(a) and 2(b) are plotted as a function of $B\cos\theta$, the $c$-axis component of the applied field.
}
\end{figure}

Figure A1 shows the data presented in Figs. 2(a) and 2(b) as a function of $B\cos\theta$.
All the curves roughly coincide, indicating that the in-plane magnetoresistance is mostly determined by the $c$-axis component of the magnetic field.

\section{Interlayer Resistivity Contaminated by In-Plane Resistivity}

\setcounter{figure}{0}
\renewcommand{\figurename}{FIG. B}

\begin{figure}
\includegraphics[width=8.6cm]{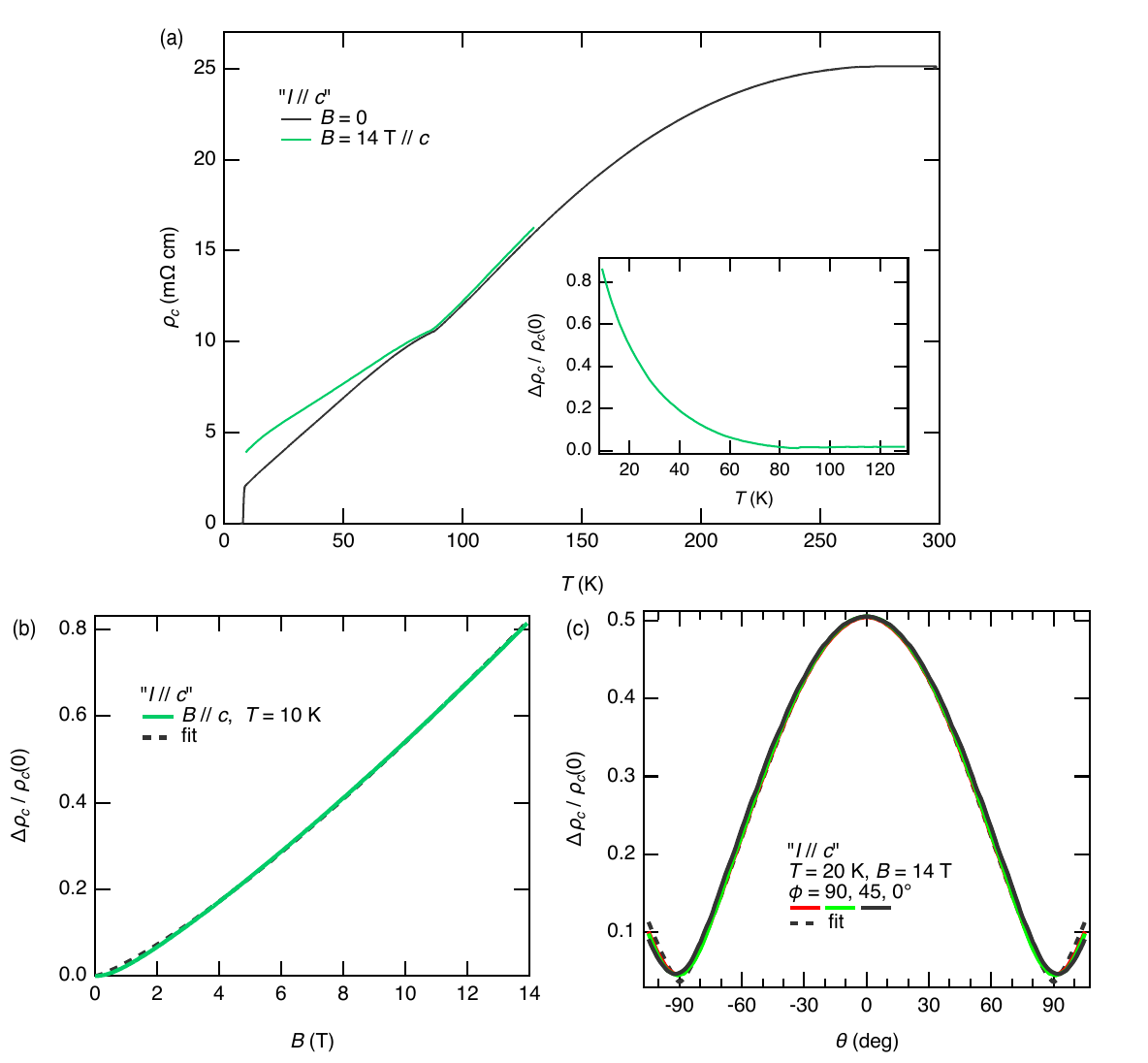}
\caption{\label{Sample}(Color online) Interlayer resistivity contaminated by in-plane resistivity in FeSe sample 3 ($0.75 \times 0.68 \times 0.08$~mm$^3$).
(a) `Interlayer resistivity' versus temperature at $B$ = 0 and 14 T applied parallel to the $c$ axis.
The inset shows the corresponding magnetoresistance.
(b) Magnetoresistance at $T$ = 10 K as a function of the magnetic field parallel to $c$.
A fit to $\alpha B^n$ (broken line) gives $\alpha$ = 0.03052(4) and $n$ = 1.2478(5).
(c) Magnetoresistance at $T$ = 20 K and $B$ = 14 T as a function of the field angle $\theta$.
Three field rotation planes $\phi$ = 0, 45, and 90$^{\circ}$ were used.
A fit to $\alpha (\cos B)^n + c$ (broken line) gives $\alpha$ = 0.470(1), $n$ = 1.335(7), and $c$ = 0.036(1).
}
\end{figure}

Figure B1 shows the results of `interlayer resistivity' measurements on sample 3.
The measurements were markedly contaminated by in-plane resistivity, as explained below.
The resistivity versus temperature curve [Fig. B1(a)] shows a broad maximum at around $T^{max} \sim$ 280 K, which is much higher than 233 K in Fig. 3.
The magnitude of the resistivity is much larger than that in Fig. 3, suggesting the occurrence of an internal cleavage.
The magnetoresistance for $B \parallel c$ is positive and large below $T_s$ down to $T_c$ (inset), similar to the in-plane magnetoresistance in Fig. 1.
Figure B1(b) shows the magnetoresistance versus magnetic field curve measured at $T$ = 10 K, which can be fitted to $\alpha B^n$ with $\alpha$ = 0.03052(4) and $n$ = 1.2478(5), similar to the in-plane magnetoresistance in Fig. 2(a), although the exponent $n$ is slightly smaller.
Figure B1(c) shows the magnetoresistance at $T$ = 20 K and $B$ = 14 T as a function of $\theta$.
Note that this field-angle dependence is unusual for the interlayer magnetoresistance in that the longitudinal magnetoresistance at $\theta$ = 0 is much larger than the transverse one at $\theta$ = $\pm$90$^{\circ}$.
The broken curve is a fit to $\alpha (B\cos \theta)^n + c$ with $\alpha$ = 0.470(1) and $n$ = 1.335(7).
The exponent $n$ is slightly different from the above value probably because of the temperature difference but is close enough, confirming that the magnetoresistance is mostly dominated by the $c$-axis component of the magnetic field as the in-plane magnetoresistance is.
A close examination of the fit near $\theta$ = $\pm$90$^{\circ}$ indicates that the experimental magnetoresistance is slightly larger than the fit, indicating a contribution from an interlayer coherence peak.

\begin{figure}
\includegraphics[width=8.6cm]{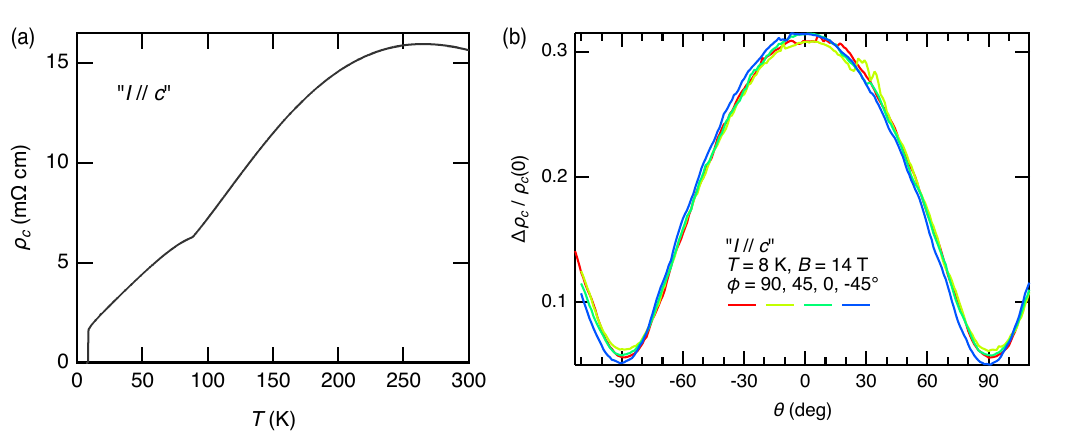}
\caption{\label{Sample}(Color online) Interlayer resistivity contaminated by in-plane resistivity in FeSe sample 4 ($1.1 \times 0.64 \times 0.24$~mm$^3$).
(a) `Interlayer resistivity' versus temperature at $B$ = 0.
(b) Magnetoresistance at $T$ = 8 K and $B$ = 14 T as a function of the field angle $\theta$.
Three field rotation planes $\phi$ = -45, 0, 45, and 90$^{\circ}$ were used.
}
\end{figure}

Figure B2 shows results of `interlayer resistivity' measurements on sample 4.
The resistivity maximum is located at $T^{max}$ = 263 K [Fig. B2(a)], which is in between the $T^{max}$ values in samples 2 and 3 (Figs. 3 and B1).
The resistivity is not as high as that in sample 3 [Figs. B2(a) and B1(a)], indicating that the internal cleavage is not so serious.
The field-angle dependence of magnetoresistance [Fig. B2(b)] is qualitatively the same as that in sample 3 [Fig. B1(c)], but the magnitude is smaller (note the temperature difference between $T$ = 8 K [Fig. B2(b)] and 20 K [Fig. B1(c)]).

In some previous papers on interlayer conductivity in layered materials, it is argued that the anomalous field-angle dependence of the interlayer magnetoresistance, like those in Figs. B1(c) and B2(b), originates from incoherent conduction along the interlayer direction. \cite{Kartsovnik09PRB, Kuraguchi03SynthMet}
In the present case, however, the clear correlation between the value of $T^{max}$ and the behavior of the apparent interlayer magnetoresistance strongly suggests that the anomalous angle dependence in Figs. B1(c) and B2(b) is due to the contamination by in-plane resistivity.
This is further corroborated by the following observations:
First, as the temperature dependence of the apparent interlayer resistivity approaches the normal metallic conduction, i.e., d$\rho$/d$T > 0$ at all temperatures, in the order of samples 2, 4, and 3 [Figs. 3(a), B2(a), and B1(a), respectively], the interlayer magnetoresistance becomes more anomalous [Figs. 4, B2(b), and B1(c)].
Secondly, when the interlayer magnetoresistance is anomalous, its field and angle dependence is described well by $(B\cos\theta)^n$ with the exponent $n$ close to that found for the in-plane magnetoresistance [Figs. B1(b) and (c)].
Note also that the quantum oscillation measurements have shown three-dimensional Fermi surfaces, i.e., modulated cylinders, not two-dimensional Fermi circles, \cite{Terashima14PRB, Audouard15EPL, Watson15PRB} which is incompatible with the incoherent scenario.

\end{document}